\documentclass[lettersize,journal]{IEEEtran}
\usepackage{amsmath,amsfonts}
\usepackage{algorithmic}
\usepackage{algorithm}
\usepackage{array}
\usepackage[caption=false,font=normalsize,labelfont=sf,textfont=sf]{subfig}
\usepackage{textcomp}
\usepackage{stfloats}
\usepackage{url}
\usepackage{pifont}
\usepackage{verbatim}
\usepackage{graphicx}
\usepackage{cite}
\usepackage{multirow}
\usepackage{balance}
\usepackage[numbers,sort&compress]{natbib}
\usepackage[colorlinks,linkcolor=blue]{hyperref}
\DeclareMathOperator{\Tr}{Tr}

\begin{document}

\title{Low-Trace Adaptation of Zero-shot Self-supervised Blind Image Denoising}

\author{Jintong Hu, 
        Bin Xia, 
        Bingchen Li, 
        Wenming Yang, ~\IEEEmembership{Senior Member,~IEEE}
\thanks{
This work was partly supported by the National Natural Science Foundation of China (Nos.62171251\&62311530100) and the Special Foundations for the Development of Strategic Emerging Industries of Shenzhen (Nos.JSGG20211108092812020\&CJGJZD20210408092804011).\\ \indent
Jintong Hu, Bingchen Li, Wenming Yang are with the Tsinghua Shenzhen International Graduate School, Tsinghua University, Shenzhen 518055, China (e-mail: hujt22@mails.tsinghua.edu.cn; libc22@mails.tsinghua.edu.cn; yang.wenming@sz.tsinghua.edu.cn). (\textit{Corresponding athor: Wenming Yang})\\ \indent
Bin Xia is with the Department of Computer Science and Engineering, The Chinese University of Hong Kong, Hongkong, China (zjbinxia@gmail.com).}
}

\markboth{Journal of \LaTeX\ Class Files}%
{Shell \MakeLowercase{\textit{et al.}}: A Sample Article Using IEEEtran.cls for IEEE Journals}


\maketitle

\begin{abstract}
Deep learning-based denoiser has been the focus of recent development on image denoising. In the past few years, there has been increasing interest in developing self-supervised denoising networks that only require noisy images, without the need for clean ground truth for training. However, a performance gap remains between current self-supervised methods and their supervised counterparts. Additionally, these methods commonly depend on assumptions about noise characteristics, thereby constraining their applicability in real-world scenarios. Inspired by the properties of the Frobenius norm expansion, we discover that incorporating a trace term reduces the optimization goal disparity between self-supervised and supervised methods, thereby enhancing the performance of self-supervised learning. To exploit this insight, we propose a trace-constraint loss function and design the low-trace adaptation Noise2Noise (LoTA-N2N) model that bridges the gap between self-supervised and supervised learning. Furthermore, we have discovered that several existing self-supervised denoising frameworks naturally fall within the proposed trace-constraint loss as subcases. Extensive experiments conducted on natural and confocal image datasets indicate that our method achieves state-of-the-art performance within the realm of zero-shot self-supervised image denoising approaches, without relying on any assumptions regarding the noise.

\end{abstract}


\section{Introduction}
\label{section 1}
\IEEEPARstart{I}{mage} denoising plays a pivotal role across various domains by addressing the issue of noise interference that can significantly compromise the quality of captured images. In critical fields such as medical diagnostics and surveillance systems, noise can conceal crucial details, posing challenges for extracting pertinent information and conducting accurate analyses. Consequently, the principal objective of image denoising is to mitigate or eliminate noise within an image, enhancing clarity and visual appeal.

\begin{figure}[h]
    \begin{minipage}[b]{1.0\linewidth}
    \centerline{\includegraphics[width=8.5cm]{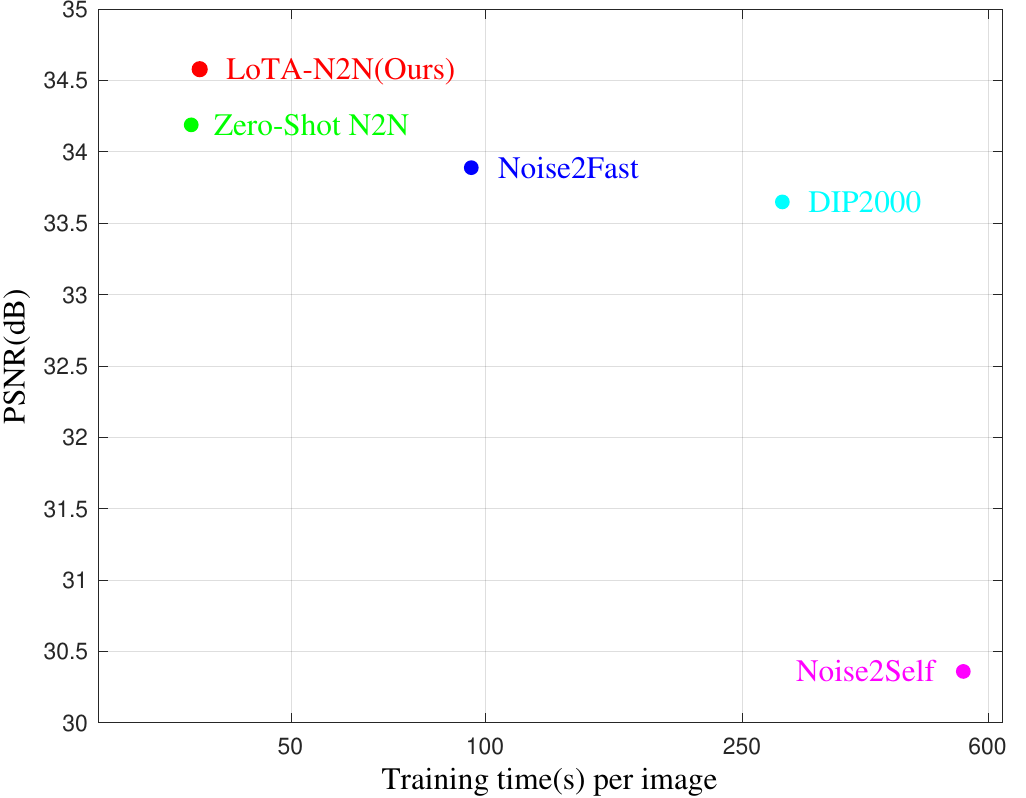}}
    
    \caption{Performance vs. training time on an RTX2080ti GPU. The results are evaluated on the McMaster18 dataset with gaussian noise $\sigma=10$. The red point represents our proposed network.}
    \label{Figure 1}
    \end{minipage}
\end{figure}

Recent advancements in deep learning have spotlighted its exceptional performance across a multitude of low-level image processing tasks \cite{dong2015image,xia2023diffir,xia2023KDSR,shi2016realtime,lim2017enhanced,ledig2017photorealistic,wang2021realesrgan,8099666,zhang2018densely,lehtinen2018noise2noise,huang2021neighbor2neighbor,7539399}. By capitalizing on extensive datasets of paired clean and noisy images, deep learning models have shown notable proficiency in noise removal, adeptly handling various noise distributions and intensities \cite{Zhang_2017,Zhang_2018,8421285,8601375,9025693,9157160,7569008,9706178,anwar2020real,guo2019convolutional}. Nonetheless, in certain spheres like biology and medical imaging, acquiring extensive clean training data can be prohibitive, both logistically and financially, if not entirely unattainable. 

Self-supervised denoising methods, which have recently aroused considerable interest and undergone extensive research, offer a novel approach to noise reduction by employing only the corrupted image, obviating the need for clean data \cite{lehtinen2018noise2noise,krull2019noise2void,wu2020unpaired,Xu_2020,Ulyanov_2020,huang2021neighbor2neighbor,lee2022apbsn,9718283,neshatavar2022cvfsid}. Contrary to supervised techniques that depend on pairs of clean and noisy images for training, these self-supervised strategies are now increasingly focused on designing lightweight models with reduced dependence on extensive training datasets \cite{10003244,10088539,9795340,mansour2023zeroshot,9853029,10052727,mansour2023zeroshot}. These advances underscore the vast potential and flexibility of self-supervised denoising in a multitude of imaging contexts. However, common assumptions about the noise characteristics, such as presumptions of a low noise level \cite{Xu_2020,9577798}, a necessity for understanding noise distribution and intensity \cite{lehtinen2018noise2noise,krull2019noise2void,9156650,9577798}, or limitations to Gaussian noise handling \cite{lehtinen2018noise2noise,9156650,soltanayev2021training}, could hinder their practical utility in complex, real-world situations.



To address the issue, we aim to bridge the gap between self-supervised and supervised denoising methods by developing loss functions that do not rely on prior noise assumptions. Our research was driven by the observation that adding a trace term to the loss function can reduce the disparity between self-supervised and supervised optimization goals, enhancing self-supervised learning performance. Through mathematical proof, we have shown that the self-supervised denoising optimization objective can be reformulated as a supervised denoising task with an added trace term, thus confirming the theoretical soundness of our approach. 

In this paper, we design the trace-constraint loss function and introduced the Low-trace Adaptation Noise2Noise model (LoTA-N2N), which proficiently enables zero-shot denoising with exceptional noise reduction capabilities. We train our LoTA-N2N in two stages: (1) Initially, during the pretraining phase, the model is trained using the mean squared error (MSE) loss. This establishes a basic but potentially biased initial denoising proficiency. At this stage, the network is trained on pairs of noisy images derived from the input noisy image, enabling it to learn to diminish noise levels without the need for clean target images. (2) Subsequently, the fine-tuning stage enhances the network's capabilities by incorporating the trace-constrained loss component. Such a strategic integration enriches the learning process, guiding the LoTA-N2N towards achieving an approximation to supervised learning paradigms without any assumptions regarding the nature of noise. Consequently, the model demonstrates a denoising capability that is robust and capable of unprecedented zero-shot noise reduction.

Our approach is anchored in the frameworks of Noise2Noise \cite{lehtinen2018noise2noise} and Neighbor2Neighbor \cite{huang2021neighbor2neighbor}, with the backbone being Zero-shot Noise2Noise \cite{mansour2023zeroshot}. This strategy ensures that our proposed model, LoTA-N2N, draws from well-established methodologies while advancing the field of denoising through innovative loss function design.
The key contributions of our LoTA-N2N model can be summarized as follows:
\begin{itemize}

    \item We introduce the trace-constraint loss, which liberates the model from reliance on any prior assumptions related to the noise model, thereby enhancing its robustness and adaptability to diverse noise distributions. This innate flexibility augments the model's practicality and efficacy across a broad spectrum of real-world scenarios.

    \item We propose LoTA-N2N, a robust, simple, and efficient zero-shot blind denoising network. Our approach employs a two-stage neural network for image denoising: it begins with MSE-based pretraining, followed by a fine-tuning phase that incorporates the trace-constrained loss, narrowing the gap between self-supervised and supervised learning and enhancing efficacy.

    \item Our model exhibits better performance and higher efficiency in image denoising. Figure \ref{Figure 1} presents the latency on an RTX2080ti GPU and PSNR of various methods. LoTA-N2N achieves the best performance and takes only 38 seconds to process a 500$\times$500 resolution image, which is 13$\%$ of the time required by DIP2000 \cite{Ulyanov_2020}.
    
\end{itemize}

\section{Related Methods}
\label{section 2}

\subsection{Theoretical Background}

Denoising refers to the process of removing noise from data, typically within the context of image processing. Noise in an image can stem from various sources, such as suboptimal lighting conditions, sensor imperfections, or transmission inconsistencies. Within the realm of deep learning, denoising involves training neural networks to discern the inherent structure of the noisy data, enabling them to predict a clean, noise-free version of the input.

Mathematically, denoising aims to approximate a function $\mathbf{f_{\theta}(\cdot)}$, parameterized by $\theta$, which maps a noisy input $\mathbf{y}$ to a corresonding clean output 
$\mathbf{x}$:
\begin{equation}
\mathbf{f_{\theta}(\mathbf{y})} \approx \mathbf{x}.
\end{equation}

Denoising methodologies can be classified into two categories based on the nature of training data: supervised and self-supervised (unsupervised). Supervised denoising requires pairs of clean and noisy data for training. The denoising function uses noisy inputs to produce denoised outputs, which are then compared to the clean data to minimize discrepancies. Such methods benefit from the direct learning signals provided by paired data, promoting a more precise understanding of the noise-to-signal mapping. In contrast, self-supervised denoising does not require labeled datasets. Instead, it aims to infer a clean data representation directly from the noisy inputs by optimizing an objective function. This function compels the network to learn the inherent structure of the data and filter out the noise. Self-supervised methods are based on the assumption that clean data reside within a lower-dimensional manifold of the noisy input space, which can be leveraged to dissociate the signal from the noise.

\subsection{Supervised Denoising Methods}

Neural networks have demonstrated significant promise in the realm of image denoising through the training of models that utilize pairs of noisy and clean images \cite{Zhang_2017,Zhang_2018,8421285,8601375,9025693,9157160,9706178,anwar2020real,guo2019convolutional}. In supervised denoising approaches, the optimization objective utilizes a loss function to train the denoising network $\mathbf{f_{\theta}(\cdot)}$, which is expressed as follows:
\begin{equation}
    \mathcal{L}_{Supervised}(\theta)=\Vert\mathbf{f_{\theta}(\mathbf{y})}-\mathbf{x}\Vert^{2}_{2},
\end{equation}
where $\mathbf{y}$ represents the noisy image, while $\mathbf{x}$ denotes its corresponding clean version. However, acquiring clean reference images in real-world scenarios is often impractical, which limits the applicability of supervised learning strategies.

The Noise2Noise (N2N) \cite{lehtinen2018noise2noise} framework addresses this limitation by replacing the clean image $\mathbf{x}$ with an independently generated noisy version $\mathbf{y'}$ from the same scene as the noisy image $\mathbf{y}$. By employing pairs of noisy images with identical static scenes, N2N attains results comparable to those obtained with noisy and clean image data pairs, provided the conditions are similar. Although procuring paired noisy images of the same scene presents practical challenges, the advent of N2N has propelled interests in sekf-supervised methods that operate on single noisy images.

\begin{figure*}
  \centering
  \includegraphics[width=1\linewidth]{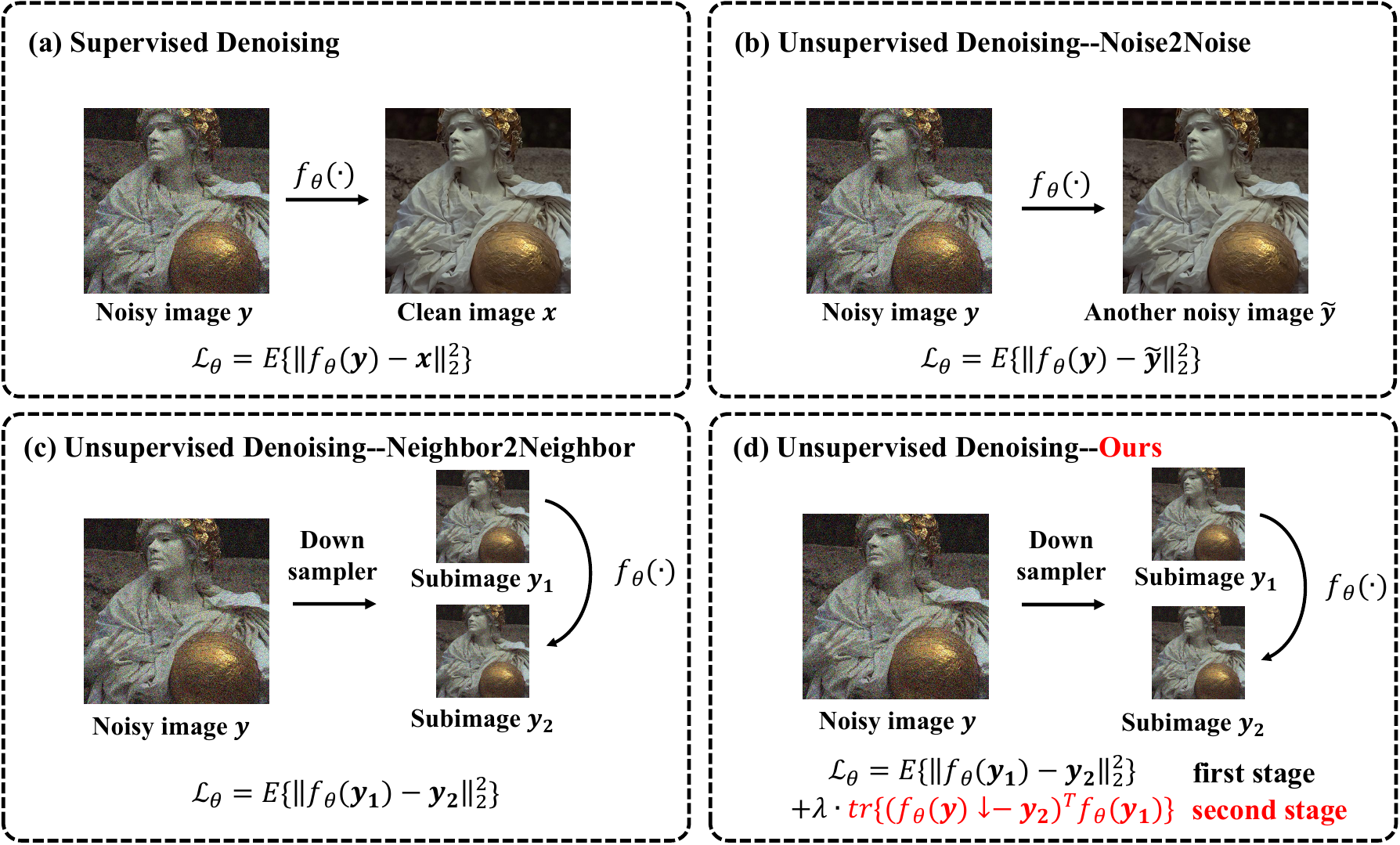}
  \caption{Comparison of different denoising methods. Supervised denoising is trained using pairs of clean/noisy images. The Noise2Noise approach circumvents the need for clean samples by employing noisy-noisy image pairs. The Neighbor2Neighbor method further refines this by generating noisy-noisy pairs through the downsampling of a single noisy image. Our method takes a further step in the loss function by constraining the trace term. It guides the self-supervised model closer to the direction of supervised learning and yields superior performance without any prior assumptions about the noise model.}
  \label{Figure 2}
\end{figure*}

\subsection{Self-Supervised Denoising Methods}

Several methods have been proposed for self-supervised image denoising in the absence of clean images \cite{lehtinen2018noise2noise,krull2019noise2void,wu2020unpaired,Xu_2020,Ulyanov_2020,huang2021neighbor2neighbor,lee2022apbsn,9157160,batson2019noise2self,neshatavar2022cvfsid,zhang2022idr,laine2019highquality}. Noise2Void (N2V) \cite{krull2019noise2void} employs blind-spot networks and modifies the N2N's loss function by replacing with the noisy image $\mathbf{y'}$ with the noisy image $\mathbf{y}$ itself. However, N2V's masking technique, designed to prevent identity mapping, leads to information loss in the masked region. Noisy as Clean (NAC) \cite{Xu_2020} makes the assumption that noise levels are minimal and demonstrates that under such conditions, the optimization objective approximates that of supervised denoising. Noisier2Noise \cite{9156650} introduces an additional noise matrix $\textbf{M}$ that follows the same distribution as the noise in the noisy image $\textbf{Y}$, generating a noisier dataset $\textbf{Z}$. The approach trains the model to map from $\textbf{Z}$ to $\textbf{Y}$ for denoising. Although NAC and Noisier2Noise provide valuable insights, their reliance on specific noise models limits their applicability to real-world scenarios where such assumptions may not hold.

Neighbor2Neighbor \cite{huang2021neighbor2neighbor} innovates by employing a neighbor-subsampling module to construct two similar sub-images, upon which the N2N training paradigm is applied. However, the resultant sub-images may not fully satisfy the N2N assumptions, posing challenges in reconciling the self-supervised and supervised learning methodologies. Iterative Denoising and Refinement (IDR) \cite{zhang2022idr} proposes a novel iterative technique to enhance the resemblance of the noisier/noisy dataset used in self-supervised learning to the noisy/clean dataset typical of supervised methods. Through this iterative refinement, IDR achieves improved denoising outcomes. Blind2Unblind \cite{9880121} circumvents the limitations of N2V by combining BSN-based results with a fully denoised image, subtly leveraging the blind-spot configuration for self-supervised training while integrating all accessible information to elevate denoising performance. Similarly, CVF-SID \cite{neshatavar2022cvfsid} deploys an array of self-supervised loss functions to segregate the clean image, independent noise map, and noise-dependent map from the input, iterating training where outputs serve as subsequent inputs to bolster component separation capabilities.

To summarize, while these pioneering techniques have advanced self-supervised denoising, they frequently rest upon assumptions about the noise characteristics that may not be valid in complex real-world contexts. This limitation often leads to suboptimal performance when these methods are applied to data with unanticipated noise distributions. Therefore, there is a clear need for denoising approaches that do not rely on any predefined assumptions about noise.

\section{Main Idea}

\label{section 3}

\subsection{Revisit of other methods}

The effectiveness of our proposed LoTA-N2N model can be theoretically supported. The discrepancy between self-supervised learning and supervised learning is attributable to their distinct optimization objectives. In our proposed method, we suggest that the loss function in self-supervised learning can be decomposed into the supervised learning loss component and an additional term. By minimizing this additional term towards zero, we can potentially align the convergence of self-supervised learning with that of supervised learning, thus achieving significant performance gains in self-supervised denoising models. To demonstrate this decomposition, we introduce the following lemmas.

\noindent
\textbf{Lemma 1.} Given a matrix $\mathbf{A}\in\mathbb{R}^{n \times n}$, the following identity holds:
\begin{equation}
\Vert\mathbf{A}\Vert^{2}_{2}=\Tr(\mathbf{A}^\text{T}\mathbf{A}),
\end{equation}
where $\Vert \cdot \Vert^{2}_{2}$ denotes the Frobenius norm (element-wise 2-norm), summed across all squared elements of the matrix, and $\Tr(\cdot)$ is the trace operation of a matrix.
\\

\noindent
\textbf{Lemma 2.} For any two matrices $\mathbf{A}$, $\mathbf{B}\in\mathbb{R}^{n \times n}$, we have:
\begin{equation}
    \Vert\mathbf{A}\pm\mathbf{B}\Vert^{2}_{2}=
    \Vert\mathbf{A}\Vert^{2}_{2}+\Vert\mathbf{B}\Vert^{2}_{2}\pm
    2\Tr (\mathbf{A}^\text{T}\mathbf{B}).
\end{equation}

\noindent$\textbf{Proof.}$ Without loss of generality, we only  show the proof for the case of subtraction as follows:
\begin{equation}
\begin{aligned}
    &\quad\ \ \Vert\mathbf{A}-\mathbf{B}\Vert^{2}_{2}\\&=\Tr(\mathbf{A}-\mathbf{B})^\text{T}(\mathbf{A}-\mathbf{B})\\
    &=\Tr(\mathbf{A}^\text{T}\mathbf{A}-\mathbf{A}^\text{T}\mathbf{B}-\mathbf{B}^\text{T}\mathbf{A}+\mathbf{B}^\text{T}\mathbf{B})\\
    &=\Tr(\mathbf{A}^\text{T}\mathbf{A})-\Tr(\mathbf{A}^\text{T}\mathbf{B})-\Tr(\mathbf{B}^\text{T}\mathbf{A})+\Tr(\mathbf{B}^\text{T}\mathbf{B})\\&=\Tr(\mathbf{A}^\text{T}\mathbf{A})-\Tr(\mathbf{A}^\text{T}\mathbf{B})-\Tr(\mathbf{A}^\text{T}\mathbf{B})+\Tr(\mathbf{B}^\text{T}\mathbf{B})\\&=\Vert\mathbf{A}\Vert^{2}_{2}-2\Tr(\mathbf{A}^\text{T}\mathbf{B})+\Vert\mathbf{B}\Vert^{2}_{2}.
\end{aligned}
\end{equation}

Using Lemma 2, we can restructure the loss of self-supervised approach as the loss in supervised learning plus or minus a trace term and a constant. The disparity between the results of self-supervised and supervised learning arises primarily from the behavior of this trace term. A logical approach might involve setting this trace term to zero, thereby bridging the gap between the performance of self-supervised and supervised learning, leading to considerable improvements in performance. In light of this, we review several prominent self-supervised denoising models: \\

\noindent
\textbf{Revisit Noise2Noise:} Noise2Noise \cite{lehtinen2018noise2noise} was a pioneering approach among self-supervised denoising methods. Instead of using noisy/clean image pairs, Noise2Noise leveraged noisy/noisy image pairs with mutually independent noise. Specifically, the pairs of noisy images in Noise2Noise can be described as follows:
\begin{equation}
\begin{aligned}
    y&=x+n,\quad\ n \sim \mathcal{N}\left(\textbf{0}, \sigma^{2}_{1}\textbf{\textit{I}}\right),\\
    y'&=x+n',\quad n \sim \mathcal{N}\left(\textbf{0}, \sigma^{2}_{2}\textbf{\textit{I}}\right),
\end{aligned}
\end{equation}
where $y$ and $y'$ constitute two independent noisy representations of a clean image $x$. Utilizing Lemma 2, the optimization objective of Noise2Noise can be reformulated:
\begin{equation}
\begin{aligned}
    &\quad\ \ \mathcal{L}_{Noise2Noise}(\theta)\\
    &=\mathbb{E}_{n, n'}\{\Vert\mathbf{f_{\theta}(\mathbf{y})}-\mathbf{y'}\Vert^{2}_{2}\}=\mathbb{E}_{n, n'}\{\Vert\mathbf{f_{\theta}(\mathbf{y})}-\mathbf{x}-\mathbf{n'}\Vert^{2}_{2}\}\\
    &=\mathbb{E}_{n, n'}\{\Vert\mathbf{f_{\theta}(\mathbf{y})}-\mathbf{x}\Vert^{2}_{2}-2\Tr\{(\mathbf{f_{\theta}(\mathbf{y})}-\mathbf{x})^\text{T}\mathbf{n'}\}+\Vert\mathbf{n'}\Vert^{2}_{2}\}\\
    &=\mathbb{E}_{n, n'}\{\Vert\mathbf{f_{\theta}(\mathbf{y})}-\mathbf{x}\Vert^{2}_{2}\}-2\mathbb{E}_{n, n'}\{\Tr\{(\mathbf{f_{\theta}(\mathbf{y})})^\text{T}\mathbf{n'}\}\}+C\\
    &=\mathbb{E}_{n, n'}\{\Vert\mathbf{f_{\theta}(\mathbf{y})}-\mathbf{x}\Vert^{2}_{2}\}-2\mathbb{E}_{n, n'}\{\Tr\{(\mathbf{n'})^\text{T}\mathbf{f_{\theta}(\mathbf{y})}\}\}+C\\
    &=\mathbb{E}_{n, n'}\{\Vert\mathbf{f_{\theta}(\mathbf{y})}-\mathbf{x}\Vert^{2}_{2}\}-2\Tr\{\mathbb{E}_{n, n'}\{(\mathbf{n'})^\text{T}\mathbf{f_{\theta}(\mathbf{y})}\}\}+C.
\end{aligned}
\end{equation}
Here, $C$ equals $\mathbb{E}_{n, n'}\{\Vert\mathbf{x}-\mathbf{y'}\Vert^{2}_{2}-2\Tr(\mathbf{x}^\text{T}(\mathbf{n'}))\}$,  which is a constant independent of $\theta$. The notation $f_{\theta}(\cdot)$ represents the denoising network characterized by learnable parameters $\theta$.

Given the statistical independence and zero-mean nature of $n$ and $n'$, we can assert:
\begin{equation}
\begin{aligned}
    &\quad \ \mathbb{E}_{n, n'}\{(\mathbf{n'})^\text{T}\mathbf{f_{\theta}(\mathbf{y})}\}\\
    &=Cov_{n, n'}((\mathbf{n'})^\text{T},\ \mathbf{f_{\theta}(\mathbf{y})})
    =Cov_{n, n'}(\boldsymbol{\sigma}\mathbf{n'}, \mathbf{M}\mathbf{y}+\mathbf{N})\\
    &=Cov_{n, n'}(\boldsymbol{\sigma}\mathbf{n'}, \mathbf{M}\mathbf{n})=\boldsymbol{\sigma}Cov\left(\mathbf{n'},\mathbf{n}\right)\mathbf{M}^\text{T}=\textbf{0}.
\end{aligned}    
\end{equation}

Accordingly, the optimization target of N2N \cite{lehtinen2018noise2noise} becomes analogous to that of supervised training, which explains why N2N achieves performance equalling or closely approaching its supervised counterparts. The proof also indicates that once $\mathbf{n}$ and $\mathbf{n}^{\prime}$ are confirmed to be mutually independent, the trace term nullifies, allowing self-supervised learning to mimic the properties of supervised learning. \\

\noindent
\textbf{Revisit Noisy As Clean:} The Noisy As Clean (NAC) \cite{Xu_2020} method posits that noise present in images is sufficiently subtle, facilitating training on a noisier/noise dataset. The method defines the noisier sample as $\mathbf{z}=\mathbf{x}+\mathbf{n}+\mathbf{m}$, and the noisy sample as $\mathbf{y}=\mathbf{x}+\mathbf{n}$, where 
$\mathbf{x}$ represents the clean image, $\mathbf{n}$ the observed noise, and $\mathbf{m}$ the simulated noise. The variances and expectations of both observed and simulated noise are presumed to be negligible. Echoing the Noise2Noise framework, the optimization objective of Noisy As Clean can be reformulated as:
\begin{equation}
\begin{aligned}
    &\quad\ \ \mathcal{L}_{Noisy As Clean}(\theta)\\
    &=\mathbb{E}_{n, m}\{\Vert\mathbf{f_{\theta}(\mathbf{z})}-\mathbf{y}\Vert^{2}_{2}\}=\mathbb{E}_{n, m}\{\Vert\mathbf{f_{\theta}(\mathbf{z})}-\mathbf{x}-\mathbf{n}\Vert^{2}_{2}\}\\
    &=\mathbb{E}_{n, m}\{\Vert\mathbf{f_{\theta}(\mathbf{y})}-\mathbf{x}\Vert^{2}_{2}\}-2\Tr\{\mathbb{E}_{n, m}\{(\mathbf{n})^\text{T}\mathbf{f_{\theta}(\mathbf{z})}\}\}+C.
\end{aligned}
\end{equation}

Here, $C$ is a constant term not dependent on $\theta$. The variables retain their meanings as defined in the previous section. Subsequently, we demonstrate that, under NAC's assumptions, the trace term is reduced to zero, illustrating how the optimization objective aligns with the supervised paradigm.
\begin{equation}
\begin{aligned}
    &\quad \ \mathbb{E}_{n, m}\{(\mathbf{n})^\text{T}\mathbf{f_{\theta}(\mathbf{z})}\}\\
    &=Cov_{n, m}((\mathbf{n})^\text{T},\ \mathbf{f_{\theta}(\mathbf{z})})+\mathbb{E}_{n, m}\{(\mathbf{n})^\text{T}\}\mathbb{E}_{n, m}\{\mathbf{f_{\theta}(\mathbf{z})}\}\\
    &\approx Cov_{n, m}((\mathbf{n})^\text{T},\ \mathbf{f_{\theta}(\mathbf{z})})=Cov_{n, m}(\boldsymbol{\sigma}\mathbf{n}, \mathbf{M}\mathbf{z}+\mathbf{N})\\
    &=Cov_{n, m}(\boldsymbol{\sigma}\mathbf{n}, \mathbf{M}\mathbf{n}+\mathbf{M}\mathbf{m})\\
    &=\boldsymbol{\sigma}Var\left(\mathbf{n}\right)\mathbf{M}^\text{T}+\boldsymbol{\sigma}Cov\left(\mathbf{n},\mathbf{m}\right)\mathbf{M}^\text{T}\\
    &\approx\boldsymbol{\sigma}\left(\rho_{n,m}\sqrt{Var(\textbf{n})}\sqrt{Var(\textbf{m})}\right)\mathbf{M}^\text{T}\\
    &\approx\textbf{0}.
\end{aligned}    
\end{equation}

\begin{figure*}
  \centering
  \includegraphics[width=\textwidth]{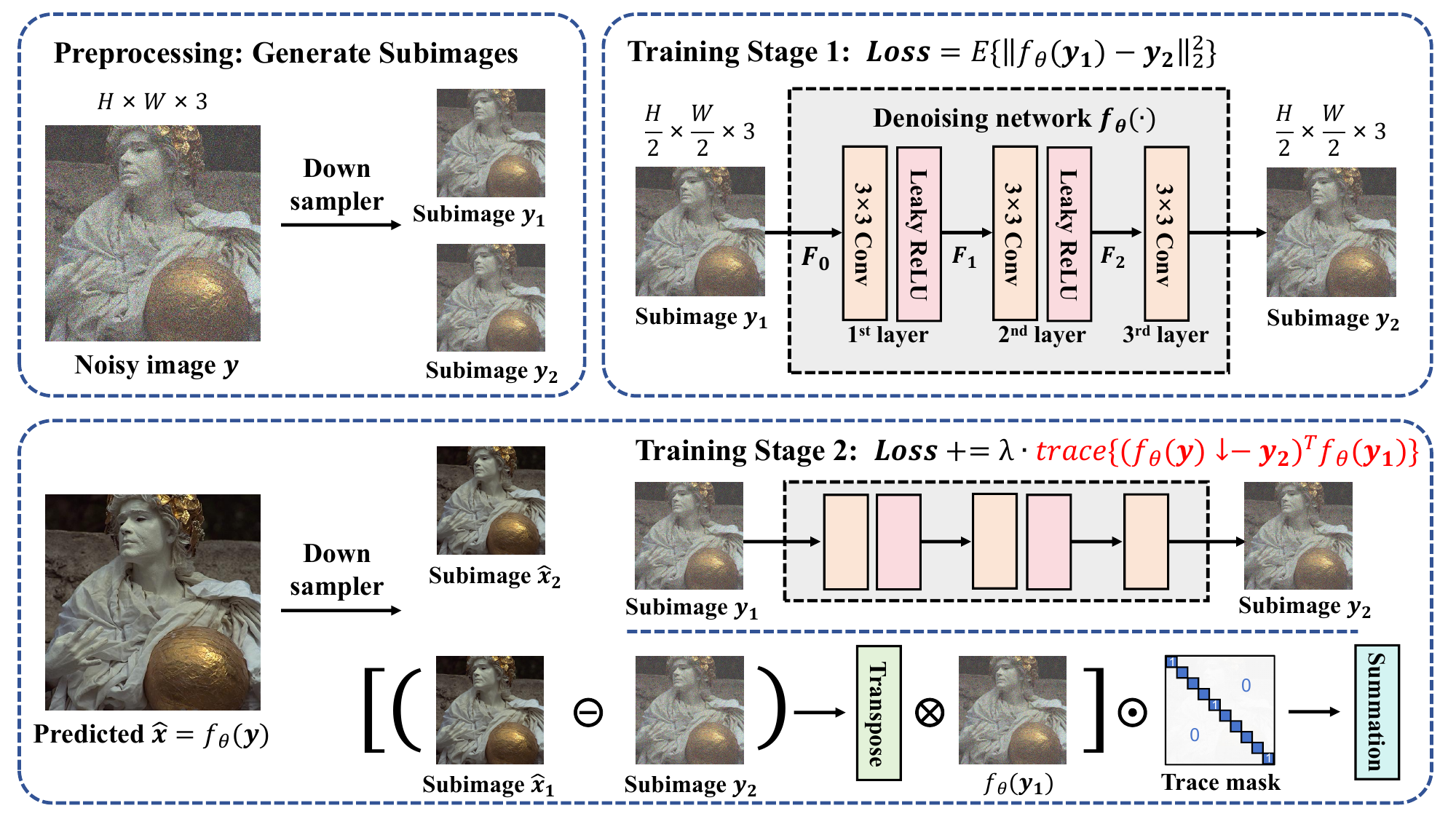}
  \caption{The main pipeline of our proposed method. The two-stage model begins with a pretraining phase where the network is initially trained using an MSE loss, leading to a biased denoiser. To improve performance, the subsequent fine-tuning stage employs the trace-constrained loss that supplements the model's training beyond the MSE baseline. This two-step training process aims to narrow the gap between self-supervised and supervised learning techniques, thus enhancing the overall effectiveness of the model.}
  \label{Figure 3}
\end{figure*}

Given this result, during the optimization process, the parameters' update direction, when applying the loss function derivative with respect to $\theta$, consistently coincides with that of a supervised learning setting. \\

\noindent
\textbf{Revisit Recorrupted2Recorrupted:} Rec2Rec \cite{9577798} generates pairs of data, $\mathbf{\widehat{y}}$ and $\mathbf{\widetilde{y}}$, both with independent noise from an initial noisy image $\mathbf{y}$. A neural network is then trained to map $\mathbf{\widehat{y}}$ to $\mathbf{\widetilde{y}}$. More formally:
\begin{equation}
    \mathbf{y}=\mathbf{x}+\mathbf{n},\quad \mathbf{n}\sim\mathcal{N}\left(\textbf{0}, \sigma^{2}\textbf{\textit{I}}\right),
\end{equation}
\begin{equation}
\mathbf{\widehat{y}}=\mathbf{y}+\mathbf{D}^\text{T}\mathbf{m},\quad\mathbf{\widetilde{y}}=\mathbf{y}-\mathbf{D}^{-1}\mathbf{m},\quad\mathbf{m}\sim\mathcal{N}\left(\textbf{0}, \sigma^{2}\textbf{\textit{I}}\right).
\end{equation}

We can establish that the trace term in the loss function of Recorrupted2Recorrupted is given by:
\begin{equation}
\begin{aligned}
    \Tr\{\mathbb{E}_{n, m}\{(\mathbf{f_{\theta}(\mathbf{\widehat{y}})})^\text{T}(\mathbf{n}-\mathbf{D}^{-1}\mathbf{m})\}\}.
\end{aligned}
\end{equation}

For simplicity, one may denote $\mathbf{\widehat{n}}=\mathbf{n}+\mathbf{D}^\text{T}\mathbf{m}$, $\mathbf{\widetilde{n}}=\mathbf{n}-\mathbf{D}^\text{T}\mathbf{m}$. The trace term can thus be rewritten as:
\begin{equation}
\begin{aligned}
    \Tr\{\mathbb{E}_{n, m}\{(\mathbf{f_{\theta}(\mathbf{x}+\mathbf{\widehat{n}})})^\text{T}\mathbf{\widetilde{n}}\}\}.
\end{aligned}
\end{equation}

Under the construction, $\mathbf{\widehat{n}}$ and $\mathbf{\widetilde{n}}$ are mutually independent, adhering to the condition discussed in the preceding Noise2Noise section. Similarly, it can be demonstrated that the trace term vanishes.




\subsection{Geometric Understanding}

In mathematics, the \textit{trace} of a matrix is defined as the sum of the elements along its main diagonal. Geometrically, this corresponds to the sum of eigenvalues of the matrix representing a linear transformation in a given coordinate system. In two dimensions, the trace encapsulates the combined scaling effects of the associated linear transformation. Thus, the trace serves as an indicator of how a transformation alters the scale of space: a positive trace signifies spatial expansion, a negative trace implies contraction, and a zero trace conveys that the size of space remains unaffected.

Consider the Noise2Noise model, where researchers set a specific matrix $(\mathbf{n'})^\text{T}\mathbf{f_{\theta}(\mathbf{y})}$ to zero based on certain noise assumptions, presupposing that the features are invariant under spatial transformations. In contrast, our proposed method requires only the trace of this matrix to be zero, which allows for the displacement of features within the space as long as such movements are balanced and the overall spatial scale is preserved. The modification substantially diminishes the dependence on noise-related assumptions and confers an appreciable advantage. Furthermore, because the trace is a scalar, integrating it into the loss function is both simpler and more efficient than setting the entire matrix to zero.

\section{Model Architecture}
\label{section 4}

Figure~\ref{Figure 2} illustrates the workflow of mainstream denoising algorithms in comparison to the process diagram of our proposed LoTA-N2N (Low-trace Adaptation Noise2Noise) model. Supervised denoising is trained using pairs of clean and noisy images. The Noise2Noise approach replaces these pairs with noisy/noisy image pairs, achieving denoising without the need for clean samples. Neighbor2Neighbor further reduces dataset requirements by generating noisy/noisy image pairs through downsampling a single noisy image. Our method, LoTA-N2N, takes a further step in the loss function by constraining the trace term. It guides the self-supervised model closer to the direction of supervised learning and yields superior performance without necessitating any prior assumptions about the noise characteristics.

In this work, we address the inherent shortcomings of conventional self-supervised denoising models that utilize the Noise2Noise (N2N) framework \cite{lehtinen2018noise2noise}, which relies on mean squared error (MSE) loss for training. Since the noisy sub-images produced by the downsampling process do not conform to the assumption of equal mean intensities, directly applying the MSE loss leads to biased estimates in the trained models. To overcome this challenge, we propose a decomposition of the MSE loss, as detailed in Section \ref{section 3}, dividing it into terms suitable for a supervised learning framework, plus an additional trace component. This methodology is expected to improve the performance of denoising networks by providing an effective strategy for more precise noise reduction in practical applications.

Initially, we use a downsampling module to split a noisy image into two similar noisy sub-images, creating the pairs required for the N2N paradigm. Let $\mathbf{y}$ denote the noisy image and the input to the downsampling module, the noisy sub-images $\mathbf{y}_{1}$ and $\mathbf{y}_{2}$ are generated as follows:
\begin{equation}
    \mathbf{y}_{1}=\mathbf{k}_{1} \otimes \mathbf{y},\quad \mathbf{y}_{2}=\mathbf{k}_{2} \otimes \mathbf{y},
\end{equation}
where $\mathbf{k}_{1}$ and $\mathbf{k}_{2}$ are two 2$\times$2 convolution kernels, and $\otimes$ denotes the convolution operation. 

As discussed in Section \ref{section 3}, the MSE loss can be decomposed into the following expression:
\begin{equation}
\begin{aligned}
    &\quad \ \mathcal{L}_{MSE}(\theta)\\
    &=\mathbb{E}\{\Vert\mathbf{f_{\theta}(\mathbf{y}_{1})}-\mathbf{y}_{2}\Vert^{2}_{2}\}
    =\mathbb{E}\{\Vert\mathbf{f_{\theta}(\mathbf{y}_{1})}-\mathbf{x}_{1}+\mathbf{x}_{1}-\mathbf{y}_{2}\Vert^{2}_{2}\}\\
    &=\mathbb{E}\{\Vert\mathbf{f_{\theta}(\mathbf{y}_1)}-\mathbf{x}_1\Vert^{2}_{2}\}+2\Tr\{\mathbb{E}\{(\mathbf{x}_{1}-\mathbf{y}_{2})^\text{T} f_{\theta}(\mathbf{y}_1)\}\}+C,
\end{aligned}
\end{equation}
where $C$ is a constant that is irrelevant to optimization and, consequently, can be discarded during the optimization process. The disparity between the optimization objectives of self-supervised and supervised denoising is thus reduced to the trace term in the equation. However, in the self-supervised denoising process, the absence of clean images leads to an inability to determine the variable $\mathbf{x}_{1}$ with precision, necessitating an estimation instead. To address this, we have designed a two-stage network architecture, employing a pre-training plus fine-tuning approach, as illustrated in Figure~\ref{Figure 3}. During the pre-training phase, we use the MSE loss to provide the model with basic denoising ability, allowing for a more accurate estimation of $\mathbf{x}_{1}$. The estimated value of $\mathbf{x}_{1}$ is given by:
\begin{equation}
    \widehat{\mathbf{x}}_{1}=\mathbf{k}_{1}\otimes f_{\theta}(\mathbf{y}).
\end{equation}
This estimated value is substituted for the true variable, yielding an approximation for the trace term, which is then integrated into the mean squared error loss, resulting in a new loss function for use in the fine-tuning training phase:
\begin{equation}
\begin{aligned}
    &\mathcal{L}_{TrC}(\theta)=\lvert\Tr\{E\{(\mathbf{x}_{1}-\mathbf{y}_{2})^\text{T}(f_{\theta}(\mathbf{y}_1)-\mathbf{x}_{1})\}\}\rvert,
\end{aligned}
\end{equation} 
\begin{equation}
\begin{aligned}
    &\mathcal{L}_{Fine-tuning}(\theta)=\mathcal{L}_{MSE}(\theta)
    +\lambda\cdot\mathcal{L}_{TrC}(\theta),
\end{aligned}
\end{equation} 
where $\mathcal{L}_{Fine-tuning}(\theta)$ is the trace-constrained loss, and $\lambda$ is a weighting factor which is subject to cosine annealing.

To improve the generalizability and robustness of the model, we incorporate the concept of mutual learning into the trace-constrained loss. This design captures transitions between noisy sub-images and imposes constraints on the reverse process, establishing a bidirectional constraint mechanism. The approach ensures that the model not only focuses on noise removal but also maintains the original structure of the image during denoising, thus enhancing the reconstruction quality. The mutual form of the trace-constrained loss is defined as:
\begin{equation}
\begin{aligned}
    \mathcal{L}_{TrC}(\theta)&=\frac{1}{2}\lvert\Tr\{E\{(\mathbf{x}_{1}-\mathbf{y}_{2})^\text{T}(f_{\theta}(\mathbf{y}_1)-\mathbf{x}_{1})\}\}\rvert\\
    &+\frac{1}{2}\lvert\Tr\{E\{(\mathbf{x}_{2}-\mathbf{y}_{1})^\text{T}(f_{\theta}(\mathbf{y}_{2})-\mathbf{x}_{2})\}\}\rvert.
\end{aligned}
\end{equation} 
\begin{table*}[!ht]
    \centering
    \setlength{\abovecaptionskip}{0pt}
    \setlength{\belowcaptionskip}{8pt}
    \footnotesize
    \setlength{\tabcolsep}{9pt}
    \renewcommand{\arraystretch}{1.15}
    \caption{Comparison of PSNR Results for Different Denoising Methods on Kodak24 and McMaster18.}
    \label{Table 1}
    \vspace{6pt}
    \begin{tabular}{|c|c|c|c|c|c|c|c|c|}
    \hline
        \multirow{2}{*}{Noise} & \multirow{2}{*}{Method} & Latency & \multicolumn{3}{c|}{\multirow{2}{*}{Kodak24: Resolution 500$\times$500}} & \multicolumn{3}{c|}{\multirow{2}{*}{McMaster18: Resolution 500$\times$500}} \\
        ~ & ~ & (s/image) & \multicolumn{3}{c|}{~} & \multicolumn{3}{c|}{~} \\
        \hline \hline
        \multirow{9}{*}{Gaussian} & ~ & ~ & $\sigma=10$ & $\sigma=15$ & $\sigma=20$ & $\sigma=10$ & $\sigma=15$ & $\sigma=20$ \\ \cline{4-9}
        ~ & \multirow{1}{*}{DnCNN \cite{Zhang_2017}} & - & 31.52 & 30.14 & 28.89  & 30.98 & 29.90 & 28.78 \\
        ~ & \multirow{1}{*}{N2N \cite{lehtinen2018noise2noise}} & - & 31.46 & 30.76 & 29.95 & 30.81 & 30.32 & 29.74 \\ \cline{2-9}
        ~ & \multirow{1}{*}{CBM3D \cite{bm3d}} & 10 & 33.50 & 31.30 & 29.83 & 34.49 & 32.18 & 30.48 \\
        ~ & \multirow{1}{*}{DIP2000 \cite{Ulyanov_2020}} & 288 & 33.13 & 31.13 & 29.69 & 33.65 & 31.86 & 30.34 \\
        ~ & \multirow{1}{*}{Noise2Self \cite{batson2019noise2self}} & 549 & 28.80 & 28.23 & 27.44 & 30.46 & 29.64 & 28.62 \\
        ~ & \multirow{1}{*}{Noise2Fast \cite{Lequyer_2022}} & 95 & 32.22 & 30.78 & 29.63 & 33.89 & 32.10 & \textbf{30.64} \\
        ~ & \multirow{1}{*}{ZSN2N \cite{mansour2023zeroshot}} & 35 & 33.91 & 31.98 & 30.43 & 34.19 & 32.00 & 30.31 \\
        ~ & \multirow{1}{*}{LoTA-N2N (Ours)} & 38 & \textbf{34.35} & \textbf{32.34} & \textbf{30.74} & \textbf{34.51} & \textbf{32.21} & 30.53 \\
        \hline \hline
        \multirow{9}{*}{Poisson} & ~ & ~ & $\lambda=60$ & $\lambda=50$ & $\lambda=40$ & $\lambda=60$ & $\lambda=50$ & $\lambda=40$ \\ \cline{4-9}
        ~ & \multirow{1}{*}{DnCNN} & - & 28.46 & 28.00 & 27.41 &  29.12 & 28.72 & 28.17 \\
        ~ & \multirow{1}{*}{N2N} & - & 29.66 & 29.31 & 28.79 & 29.68 & 29.43 & 29.03 \\ \cline{2-9}
        ~ & \multirow{1}{*}{CBM3D} & 10 & 28.33 & 28.26 & 28.08 & 29.33 & 29.21 & 28.97 \\
        ~ & \multirow{1}{*}{DIP2000} & 288 & 29.11 & 28.62 & 28.04 & 30.29 & 29.78 & 29.33 \\
        ~ & \multirow{1}{*}{Noise2Self} & 549 & 27.08 & 26.77 & 26.67 & 29.03 & 29.00 & 28.31 \\
        ~ & \multirow{1}{*}{Noise2Fast} & 95 & 29.29 & 28.87 & 28.37 & 31.01 & 30.54 & 29.98 \\
        ~ & \multirow{1}{*}{ZSN2N} & 35 & 30.36 & 29.93 & 29.28 & 30.80 & 30.47 & 29.86 \\
        ~ & \multirow{1}{*}{LoTA-N2N (Ours)} & 38 & \textbf{30.54} & \textbf{30.12} & \textbf{29.52} & \textbf{31.09} & \textbf{30.69} & \textbf{30.07} \\ \hline
    \end{tabular}
\end{table*}

\begin{table}[!ht]
    \centering
    \setlength{\abovecaptionskip}{0pt}
    \setlength{\belowcaptionskip}{8pt}
    \footnotesize
    \setlength{\tabcolsep}{9pt}
    \renewcommand{\arraystretch}{1.15}
    \caption{Comparison of PSNR and SSIM Results for Different Denoising Methods on Set14 and BSD68.} 
    \label{Table 2}
    \vspace{6pt}
    \begin{tabular}{|c|c|c|c|c|}
    \hline
        \multicolumn{1}{|c}{Method} & \multicolumn{4}{|c|}{Set14 (Upper) and BSD68 (Below)} \\ 
        \hline \hline
        ~ & $\sigma=10$ & $\sigma=15$ & $\sigma=20$ & $\sigma=25$ \\ 
        \cline{2-5}
        \multirow{2}{*}{CBM3D} & 32.92 & 30.74 & 29.22 & 28.03 \\
        ~ & 0.9448 & 0.9156 & 0.8887 & 0.8634 \\ \hline
        \multirow{2}{*}{DIP2000} & 29.91 & 29.26 & 28.26 & 27.41 \\ 
        ~ & 0.8463 & 0.8236 & 0.7902 & 0.7652 \\ \hline
        \multirow{2}{*}{Noise2Fast} & 31.49 & 30.08 & 28.93 & 27.94 \\ 
        ~ & 0.8707 & 0.8351 & 0.8037 & 0.7733 \\ \hline
        \multirow{2}{*}{ZSN2N} & 32.90 & 31.00 & 29.51 & 28.25 \\
        ~ & 0.9446 & 0.9217 & 0.8964 & 0.8715 \\ \hline
        \multirow{2}{*}{LoTA-N2N (Ours)} & \textbf{32.96} & \textbf{31.10} & \textbf{29.62} & \textbf{28.46} \\ 
        ~ & \textbf{0.9464} & \textbf{0.9243} & \textbf{0.8988} & \textbf{0.8753} \\
        \hline \hline
        ~ & $\sigma=10$ & $\sigma=15$ & $\sigma=20$ & $\sigma=25$ \\  \cline{2-5}
        \multirow{2}{*}{CBM3D} & 32.70 & 30.35 & 28.80 & 27.65 \\
        ~ & 0.9485 & 0.9166 & 0.8871 & 0.8601 \\ \hline
        \multirow{2}{*}{DIP2000} & 31.36 & 30.49 & 29.41 & 28.28 \\
        ~ & 0.9284 & 0.9162 & 0.8986 & 0.8814 \\ \hline
        \multirow{2}{*}{Noise2Fast} & 31.36 & 29.91 & 28.77 & 27.83 \\
        ~ & 0.8901 & 0.8519 & 0.8167 & 0.7853 \\ \hline
        \multirow{2}{*}{ZSN2N} & 34.62 & 32.31 & 30.61 & 29.26 \\
        ~ & 0.9651 & 0.9428 & 0.9190 & 0.8950 \\ \hline
        \multirow{2}{*}{LoTA-N2N (Ours)} & \textbf{34.64} & \textbf{32.46} & \textbf{30.73} & \textbf{29.36} \\ 
        ~ & \textbf{0.9686} & \textbf{0.9484} & \textbf{0.9242} & \textbf{0.8989} \\ \hline
    \end{tabular}
\end{table}

Further, we enhance the trace-constrained loss function by incorporating principles of residual learning, which posits that separating noise is less challenging than reconstructing an uncorrupted image given that the noise typically exhibits lower amplitudes and variance compared to the signal. These properties facilitate a more precise noise estimation. Subsequently, we refine our algorithm to focus on extracting the noise component-the residual-rather than attempting to reproduce the pristine image. The residual enhancement is quantified by:
\begin{equation}
\begin{aligned}
    \widehat{\mathbf{x}}=\mathbf{y}-f_{\theta}(\mathbf{y}).
\end{aligned}
\end{equation}

In summary, our two-stage neural network approach for image denoising begins with a pretraining phase where the network is initially trained using an MSE loss function, leading to a biased denoiser. To improve performance, the subsequent fine-tuning stage employs additional loss components inspired by mutual and residual learning concepts, resulting in a trace-constrained loss that supplements the model's training beyond the MSE baseline. The two-step training process aims to narrow the gap between self-supervised and supervised learning, thus improving the overall effectiveness of the model.

\section{Experiment Results}
\label{section 5}

\subsection{Datasets and Evaluation Metrics}

To evaluate the effectiveness of our algorithm, we conducted experiments using four natural image datasets: Kodak24\footnote{ source: https://r0k.us/graphics/kodak}, McMaster18 \cite{McMaster18}, Set14 \cite{Set14}, and BSD68 \cite{BSD68}. The Kodak24 dataset consists of 24 color natural images with a resolution of 500$\times$500 pixels, while the McMaster18 dataset includes 18 color natural images of the same resolution. Set14 comprises a diverse collection of 14 images, each varying in size from dimensions as large as 768×512 to as compact as 276×276 pixels, featuring a variety of natural scenes and artificial objects. The BSD68 dataset, derived from the larger Berkeley Segmentation Dataset, contains 68 high-quality, clear images with native dimensions of 481$\times$321 pixels. Each dataset is extensively utilized in the literature for benchmarking state-of-the-art image processing algorithms. Additionally, to further assess the performance of our method, we included confocal and medical imaging data in our evaluation.

The confocal data used in our study was obtained from the Fluorescent Microscopy Dataset (FMD) \cite{zhang2019poissongaussian}, which provides a collection of high-resolution images critical for biological specimen analysis. To ensure a fair comparison, we used the same image samples as those employed by the Noise2Fast method, maintaining consistency in our benchmarking approach. The medical imaging data were sourced from pediatric chest X-ray images specified in \cite{xray}, comprising 5,232 images from 5,856 patients. Within the dataset, 3,883 images showcased pneumonia, with 2,538 due to bacterial and 1,345 due to viral infections. Additionally, 1,349 images of normal chest X-rays were included for control. We randomly selected a subset of 17 normal chest X-ray images as our training set, which were then resized to a resolution of 800$\times$800 pixels through random cropping. 

In accordance with prior research, our primary evaluation metrics are the peak signal-to-noise ratio (PSNR) and the structural similarity index measure (SSIM) \cite{ssim}.

\begin{table}[!ht]
    \centering
    
    \setlength{\abovecaptionskip}{0pt}
    \setlength{\belowcaptionskip}{8pt}
    \footnotesize
    \setlength{\tabcolsep}{9pt}
    \renewcommand{\arraystretch}{1.15}
    \caption{Comparison of PSNR Results for Different Denoising Methods on Confocal and MRI Dataset.} 
    \label{Table 3}
    \vspace{6pt}
    \begin{tabular}{|c|c|c|c|c|}
    \hline
        \multicolumn{1}{|c}{Method} & \multicolumn{4}{|c|}{Confocal: Resolution 500$\times$500} \\ 
        \hline \hline
        ~ & $\sigma=5$ & $\sigma=10$ & $\lambda=60$ & $\lambda=50$ \\ 
        \cline{2-5}
        \multirow{1}{*}{CBM3D} & 42.47 & 38.28 & 36.87 & 36.70 \\
        \multirow{1}{*}{DIP2000} & 38.98 & 37.11 & 37.20 & 36.84 \\
        \multirow{1}{*}{Noise2Fast} & 41.49 & 38.98 & 38.52 & 38.25 \\
        \multirow{1}{*}{ZSN2N} & 44.13 & 39.01 & 39.81 & 39.33 \\
        \multirow{1}{*}{LoTA-N2N (Ours)} & \textbf{44.21} & \textbf{39.26} & \textbf{40.17} & \textbf{39.65} \\
        \hline \hline
        
        \multicolumn{1}{|c}{Method} & \multicolumn{4}{|c|}{X-Ray: Resolution 800$\times$800} \\ 
        \hline \hline
        ~ & $\sigma=5$ & $\sigma=10$ & $\lambda=60$ & $\lambda=50$ \\ 
        \cline{2-5}
        \multirow{1}{*}{CBM3D} & 41.30 & 38.60 & 35.57 & 35.25 \\
        \multirow{1}{*}{DIP2000} & 36.21 & 35.95 & 35.53 & 35.33 \\
        \multirow{1}{*}{Noise2Fast} & 40.83 & 38.40 & 35.32 & 34.84 \\
        \multirow{1}{*}{ZSN2N} & 42.04 & 39.06 & 35.79 & 35.31 \\
        \multirow{1}{*}{LoTA-N2N (Ours)} & \textbf{42.96} & \textbf{39.74} & \textbf{35.81} & \textbf{35.35} \\
        \hline
    \end{tabular}
\end{table}

\begin{figure*}
  \centering
  \includegraphics[width=\textwidth]{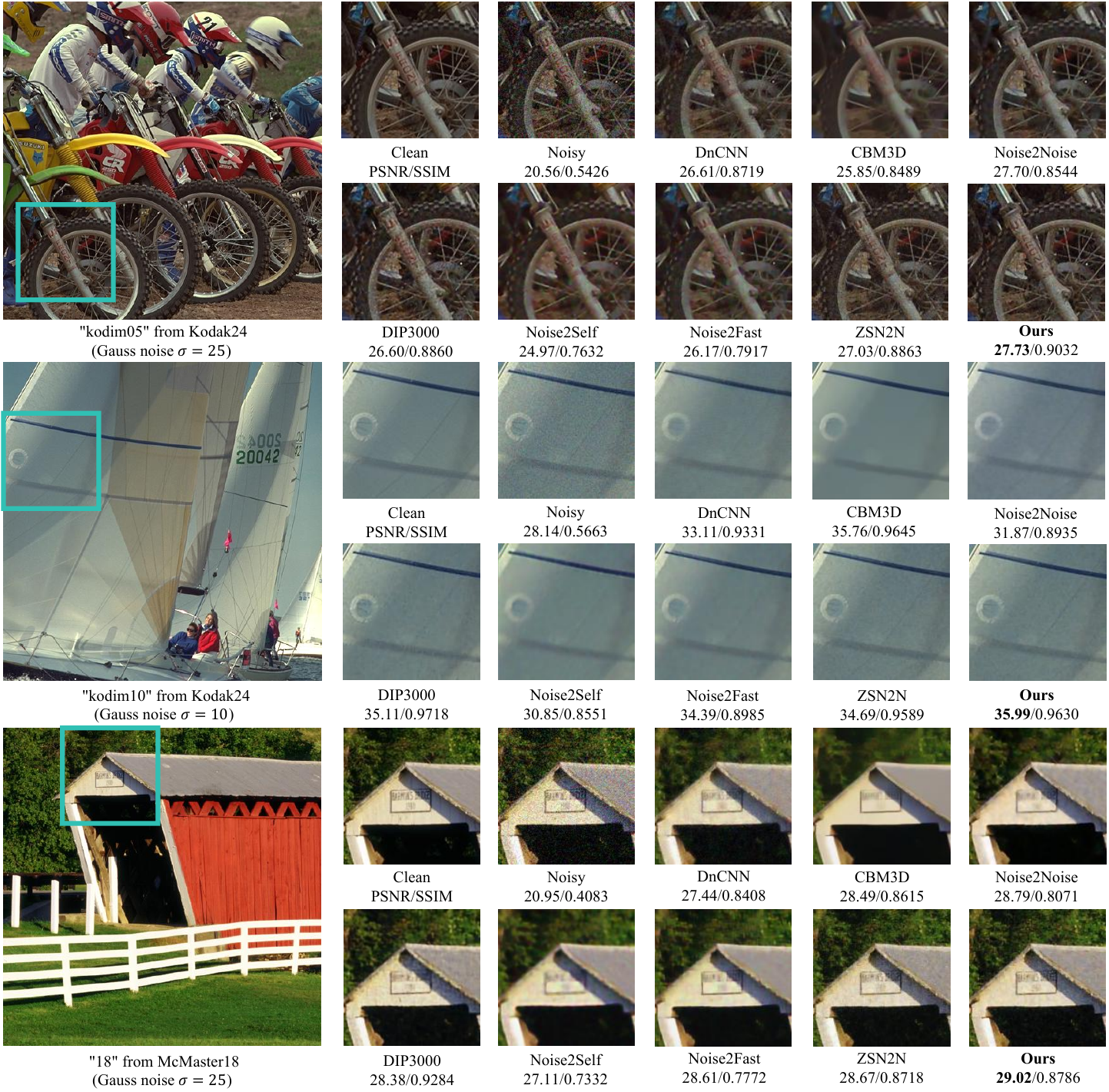}
  \caption{Visual comparison between methods. Our proposed denoising approach demonstrates superior performance in preserving the fidelity of textural details, particularly in texture-rich regions, achieving the best denoising results compared to other methods.
  }
  \label{Figure 4}
\end{figure*}

\subsection{Comparison with other methods}

We trained and tested our model specifically for Gaussian and Poisson noise levels. Poisson noise, also known as shot noise, is a type of noise in which the pixel values vary according to a Poisson distribution contingent on the intensity of the underlying signal. In contrast to additive noise, which introduces a constant or scaled noise value to the signal, Poisson noise is signal-dependent-regions with higher intensity in an image will manifest a greater amount of noise.

In our experiments, we employed the CBM3D variant of the BM3D method \cite{bm3d} to perform noise reduction on multichannel images. The CBM3D method, when trained on Gaussian noise with known noise variances, displayed performance that sometimes surpassed even the most recent methodologies. Models such as DnCNN \cite{Zhang_2017} and Noise2Noise \cite{lehtinen2018noise2noise} were trained and tested entirely on the same dataset. For the Noise2Noise model, we conducted training over 100 epochs on the Kodak24 dataset, and extended the training to 250 epochs on the McMaster18 dataset. With the Deep Image Prior (DIP) model \cite{Ulyanov_2020}, our empirical findings indicated that convergence occurred without further gains beyond 1500 epochs; in fact, proceeding past this threshold led to deteriorating results. As such, for DIP, we capped the maximum number of epochs at 2000, henceforth referred to as DIP2000. We utilized the single-shot version of the noise2self framework \cite{batson2019noise2self}, as provided by its authors. For all other models under consideration, their default parameter settings were retained.

\begin{table*}[!h] 
    \label{Ablation study}
    \centering
    \setlength{\abovecaptionskip}{0pt}
    \setlength{\belowcaptionskip}{8pt}
    \footnotesize
    \setlength{\tabcolsep}{9pt}
    \renewcommand{\arraystretch}{1.15}
    \caption{Ablation study of LoTA-N2N on trace-constraint loss(TrCL), residual enhancement and mutual learning. The evaluation is performed on the McMaster18, with a focus on measuring the peak signal-to-noise ratio (PSNR) and structure similarity index measure (SSIM) to assess the performance of these strategies. The best result of each noise level is in bold.}
    \label{Table 4}
    \vspace{6pt}
    \begin{tabular}{|c|c|c|c|c c c c|}
    \hline
        \multirow{2}{*}{Setting} & \multirow{2}{*}{TrCL} & \multirow{2}{*}{Mutual learning} & \multirow{2}{*}{Residual enhancement} & \multicolumn{4}{c|}{Noise level (PSNR $\uparrow$ / SSIM $\uparrow$)} \\ \cline{5-8}
        ~ & ~ & ~ & ~ & $\sigma=10$ & $\sigma=15$ & $\sigma=20$ & $\sigma=25$ \\ \hline
        $S_{1}$ & \ding{56} & \ding{56} & \ding{56} & 33.70 / 0.9457 & 31.66 / 0.9126 & 30.07 / 0.8804 & 28.80 / 0.8501 \\ 
        $S_{2}$ & \ding{52} & \ding{56} & \ding{56} & 33.78 / 0.9461 & 31.76 / 0.9153 & 30.17 / 0.8844 & 28.86 / 0.8537 \\ 
        $S_{3}$ & \ding{52} & \ding{52} & \ding{56} & 34.05 / 0.9471 & 31.90 / 0.9159 & 30.40 / 0.8868 & 29.03 / 0.8568 \\ 
        $S_{4}$ & \ding{52} & \ding{52} & \ding{52} & \textbf{34.51} / \textbf{0.9539} & \textbf{32.21} / \textbf{0.9251} & \textbf{30.53} / \textbf{0.8922} & \textbf{29.11} / \textbf{0.8593} \\ 
        \hline
    \end{tabular}
\end{table*}

\begin{figure}[h]
    \begin{minipage}[b]{1.0\linewidth}
    \centerline{\includegraphics[width=8.5cm]{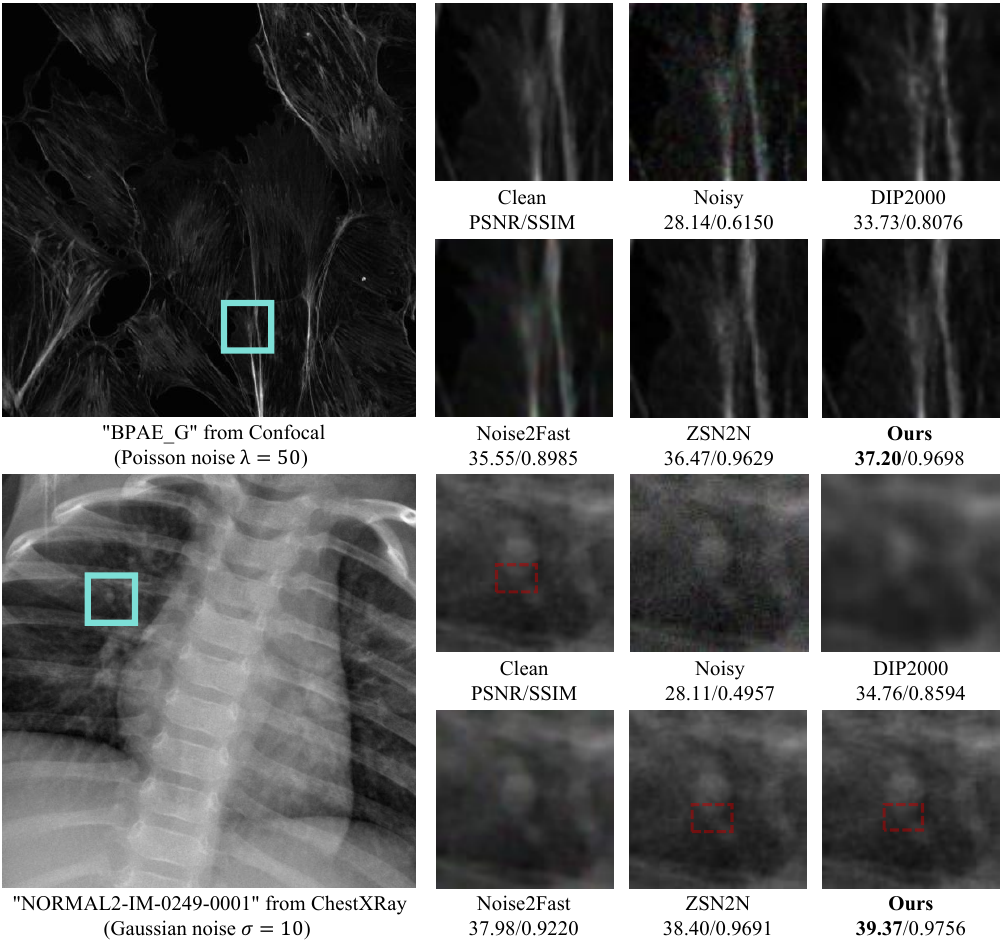}}
    \caption{Visual comparison on confocal and medical datasets. Our approach maintains a greater level of detail within regions abundant in texture.}
    \vspace{-0.2cm}
    \label{Figure 5}
    \end{minipage}
\end{figure}

\begin{figure}[h]
    \begin{minipage}[b]{1.0\linewidth}
        \centerline{\includegraphics[width=8.5cm]{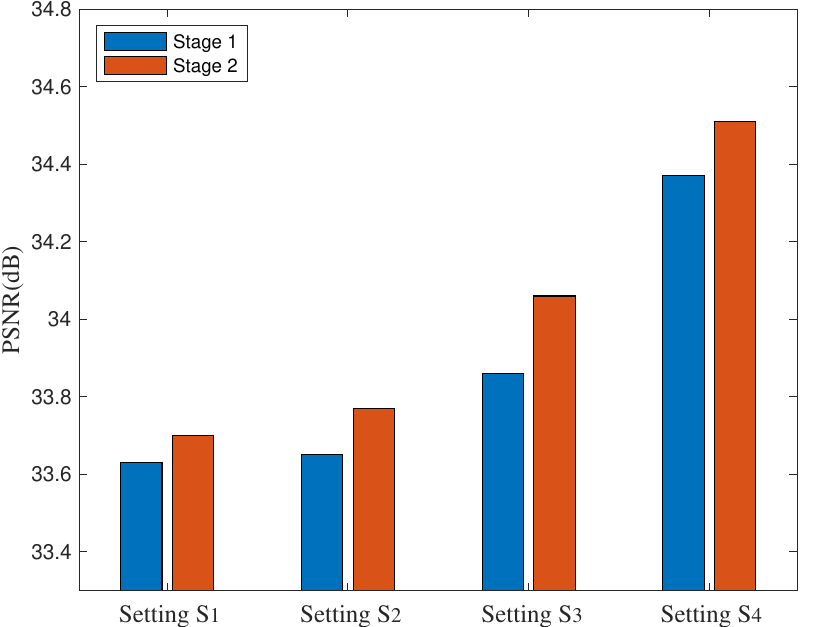}}
        \caption{Comparative Analysis of PSNR (dB) Outcomes for Two-Phase Training Across Different Model Configurations ($S_{1}$-$S_{4}$).}
        \label{Figure 6}
        \vspace{-0.2cm}
    \end{minipage}
\end{figure}

Table \ref{Table 1} showcases the qualitative outcomes on natural image datasets, Kodak24 and McMaster18. Our method outstrips contemporary techniques across various noise levels while also demonstrating reduced latency on an RTX2080ti GPU. Figure~\ref{Figure 4} visually emphasizes our method's superiority in diminishing noise and preserving high-fidelity textural details. LoTA-N2N effectively restored fine text details without introducing artifacts or exhibiting jagged textures.

Table \ref{Table 2} presents the results on the Set14 and BSD68 datasets, where our LoTA-N2N consistently outperformed other methods across all evaluated noise levels. Furthermore, our methodology has shown promising results in the biomedical domain, as evidenced by the performance metrics presented in Table \ref{Table 3}, which include analyses on both confocal and X-ray datasets. In addition to its enhanced denoising capabilities, our model further distinguishes itself through significantly reduced computation time. These attributes collectively exemplify an advantageous synergy of performance efficacy and computational expediency. Additionally, Figure~\ref{Figure 5} presents a visual comparison on confocal and X-ray datasets, with the most significant differences highlighted within cyan line boxes. The upper section demonstrates the results on the confocal dataset, wherein our approach delivers the clearest detail and texture without introducing artifacts observed in approaches like Noise2Fast. Compared with the ZSN2N method, our technique preserves the finest features, particularly at the center of the display frame, demonstrating superior restoration capabilities. The lower portion illustrates the results on a pulmonary X-ray dataset. Here, the ZSN2N method unfortunately introduces spurious texture structures not present in the original image, as indicated by the red dashed-line boxes. In contrast, DIP and Noise2Fast struggle to recover such intricate texture, while our method continues to display robust denoising performance, producing images that most closely resemble the clear samples.


\subsection{Ablation Study}

To further demonstrate our model's effectiveness, we conducted an ablation study on our LoTA-N2N. This study includes an analysis of modules such as trace-constrained loss (TrCL), residual enhancement, and mutual study. 

Table \ref{Table 4} presents the detailed results of the ablation study. The baseline model, denoted as $S_{1}$, employs MSE loss for training through two distinct phases without incorporating either residual enhancement or mutual study. This baseline model was extended to include trace-constrained loss, resulting in configuration $S_{2}$. Upon introducing TrCL, enhanced performance was observed across various noise levels. Further refinements to $S_{2}$ entailed applying a mutual study paradigm to the loss function, yielding configuration $S_{3}$. This adaptation imposes constraints on both the forward and the inverse processes, leading to additional improvements in the performance metrics. In a subsequent enhancement, residual learning was incorporated, enabling the model to distinguish between clean and noisy image components by learning the characteristics of the noise. This strategy proved effective in reducing the variance of the results, which in turn increased their precision. The fully developed model, represented as $S_{4}$, incorporates the trace-constrained loss, residual enhancement, and mutual study form. This final model configuration achieved the most favorable results. The stepwise progression from $S_{1}$ to $S_{4}$ serves to confirm the validity and effectiveness of the proposed modules within the overall framework.

Further experiments were designed to assess our two-phase training strategy. The results are visually depicted in Figure~\ref{Figure 6}. In the first phase, the model is trained using MSE loss, followed by a second phase where TrCL is incorporated. Ablation studies evaluated the impact of these training stages. Under condition $S_{1}$, both network phases employed MSE loss; the resulting PSNR metrics were nearly identical with no significant differences observed, indicating that without the inclusion of TrCL, the two-phase approach does not offer measurable enhancements in terms of PSNR. For condition $S_{2}$, the introduction of TrCL during the second training phase yielded notably different results between the two stages, substantiating the efficacy of fine-tuning with TrCL. When the mutual study paradigm was applied to the loss function under condition $S_{3}$, improvements were observed in both stages; however, the second phase achieved superior results, underscoring the benefits of this design. With $S_{4}$, the implementation of both mutual study and residual enhancement resulted in great improvements across both phases. Notably, the second fine-tuning phase, which utilized TrCL, maintained a significant lead over the initial training phase, providing further corroboration of the effectiveness of the designed modules in LoTA-N2N.

\section{Conclusion}
\label{section 6}

In this paper, we propose a novel trace-constraint loss function that bridges the gap between self-supervised and supervised learning in the field of image denoising. By effectively optimizing the self-supervised denoising objective through the incorporation of a trace term as a constraint, our approach allows for improved performance and generalization across various types of images including natural, medical, and biological imagery. We enhance the designed trace-constraint loss function by incorporating the concepts of mutual study and residual study to achieve improved denoising performance and generalization. Furthermore, our designed model has been kept lightweight, enabling better denoising results to be achieved in a shorter training time, without the need for any prior assumptions about the noise. Our method outperforms existing self-supervised denoising models by a significant margin, demonstrating its potential for widespread adoption and practicality in real-world scenarios. Overall, our approach represents a valuable contribution to the advancement of self-supervised denoising methods and holds promise for addressing practical challenges associated with acquiring paired clean / noisy images for supervised learning.

\bibliographystyle{IEEEtran}
\bibliography{bare_jrnl_new_sample4}

\vfill

\end{document}